# On Predictive Modeling for Optimizing Transaction Execution in Parallel OLTP Systems

Andrew Pavlo
Brown University
pavlo@cs.brown.edu

Evan P.C. Jones
MIT CSAIL
evanj@mit.edu

Stanley Zdonik
Brown University
sbz@cs.brown.edu

## ABSTRACT

A new emerging class of parallel database management systems (DBMS) is designed to take advantage of the partitionable workloads of on-line transaction processing (OLTP) applications [23, 20]. Transactions in these systems are optimized to execute to completion on a single node in a shared-nothing cluster without needing to coordinate with other nodes or use expensive concurrency control measures [18]. But some OLTP applications cannot be partitioned such that all of their transactions execute within a single-partition in this manner. These distributed transactions access data not stored within their local partitions and subsequently require more heavy-weight concurrency control protocols. Further difficulties arise when the transaction's execution properties, such as the number of partitions it may need to access or whether it will abort, are not known beforehand. The DBMS could mitigate these performance issues if it is provided with additional information about transactions. Thus, in this paper we present a Markov model-based approach for automatically selecting which optimizations a DBMS could use, namely (1) more efficient concurrency control schemes, (2) intelligent scheduling, (3) reduced undo logging, and (4) speculative execution. To evaluate our techniques, we implemented our models and integrated them into a parallel, main-memory OLTP DBMS to show that we can improve the performance of applications with diverse workloads.

## 1. INTRODUCTION

Shared-nothing parallel databases are touted for their ability to execute OLTP workloads with high throughput. In such systems, data is spread across shared-nothing servers into disjoint segments called *partitions*. OLTP workloads have three salient characteristics that make them amenable to this environment: (1) transactions are short-lived (i.e., no user stalls), (2) transactions touch a small subset of data using index look-ups (i.e., no full table scans or large distributed joins), and (3) transactions are repetitive (i.e., executing the same queries with different inputs) [23].

Even with careful partitioning [7], achieving good performance with this architecture requires significant tuning because of distributed transactions that access multiple partitions. Such transactions require the DBMS to either (1) block other transactions from using each partition until that transaction finishes or (2) use fine-grained locking with deadlock detection to execute transactions concurrently [18]. In either strategy, the DBMS may also need to maintain an undo buffer in case the transaction aborts. Avoiding such onerous concurrency control is important, since it has been shown to be approximately 30% of the CPU overhead for OLTP workloads in traditional databases [14]. To do so, however, requires the DBMS to have additional information about transactions before they start. For example, if the DBMS knows that a transaction only needs to access data at one partition, then that transaction can be redirected to the machine with that data and executed without heavy-weight concurrency control schemes [23].

It is not practical, however, to require users to explicitly inform the DBMS how individual transactions are going to behave. This is especially true for complex applications where a change in the database's configuration, such as its partitioning scheme, affects transactions' execution properties. Hence, in this paper we present a novel method to automatically select which optimizations the DBMS can apply to transactions at runtime using Markov models. A Markov model is a probabilistic model that, given the current state of a transaction (e.g., which query it just executed), captures the probability distribution of what actions that transaction will perform in the future. Based on this prediction, the DBMS can then enable the proper optimizations. Our approach has minimal overhead, and thus it can be used on-line to observe requests to make immediate predictions on transaction behavior without additional information from the user. We assume that the benefit outweighs the cost when the prediction is wrong. This paper is focused on stored procedure-based transactions, which have four properties that can be exploited if they are known in advance: (1) how much data is accessed on each node, (2) what partitions will the transaction read/write, (3) whether the transaction could abort, and (4) when the transaction will be finished with a partition.

We begin with an overview of the optimizations used to improve the throughput of OLTP workloads. We then describe our primary contribution: representing transactions as Markov models in a way that allows a DBMS to decide which of these optimizations to employ based on the most likely behavior of a transaction. Next, we present *Houdini*, an on-line framework that uses these models to generate predictions about transactions before they start. We have integrated this framework into the H-Store system [2] and measure its ability to optimize three OLTP benchmarks. The results from these experiments demonstrate that our models select the proper optimizations for 93% of transactions and improve the throughput of the system by 41% on average with an overhead of 5% of the total transaction execution time. Although our work is described in the context of H-Store, it is applicable to similar OLTP systems.

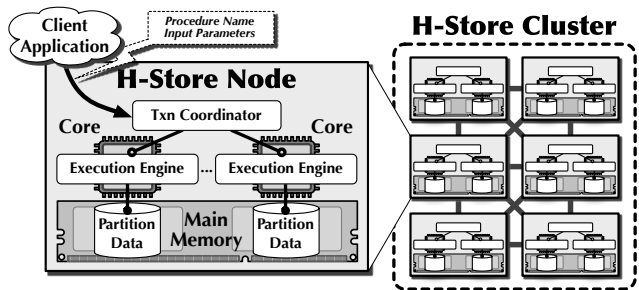

**Figure 1:** The H-Store Main Memory OLTP system.

## 2. TRANSACTION OPTIMIZATIONS

We first discuss the optimizations that are possible if one knows what a transaction will do prior to its execution in a stored procedure-based DBMS. Stored procedures are an effective way to optimize OLTP applications because they reduce the number of round-trips between the client and the database, thereby eliminating most network overhead and shrinking the window for lock contention. They contain parameterized queries separated by *control code* (i.e., application logic), and thus most DBMSs do not know what each transaction invocation of a procedure will do at run time (e.g., what set of pre-defined queries it will execute and what partitions those queries will access). This is because the procedure can contain loops and conditionals that depend on the parameters from the application and the current values stored in the database.

One example of this type of system is H-Store, a parallel, row-storage relational OLTP DBMS that runs on a cluster of shared-nothing, main memory-only nodes [23, 20]. It is currently being developed by Brown, M.I.T., and Yale [2]. A commercial version, called VoltDB, is also being developed based on its design. As shown in Fig. 1, an H-Store node consists of a transaction coordinator that manages single-threaded execution engines, each with exclusive access to a data partition stored in memory. Client applications initiate transactions by sending the pre-defined procedure name and input to any node in the cluster.

We now discuss the four transaction optimizations that OLTP systems like H-Store can employ at run time if they know certain properties about transactions before they begin to execute.

**OP1. Execute the transaction at the node with the partition that it will access the most.**

When a new transaction request is received, the DBMS's transaction coordinator must determine which node in the cluster should execute the procedure's control code and dispatch queries. In most systems, this node also manages a partition of data. We call this the *base partition* for a transaction. The "best" base partition is the one containing most of the data that will be accessed by that transaction, as that reduces the amount of data movement. Any transaction that needs to access only one data partition is known as a *single-partition* transaction. These transactions can be executed efficiently on a parallel DBMS, as they do not require multi-node coordination [23]. Hence, determining the correct base partition will dramatically increase throughput and decrease latency in any distributed database that supports stored procedures.

One naïve strategy is to execute each transaction on a random partition to evenly distribute work, but the likelihood that this approach picks the "wrong" partition increases with the number of partitions. An alternative approach, used by IBM's DB2, is to execute the procedure on any node, then if the first statement accesses data in some other partition, abort and re-start the transaction there [6]. This heuristic does not work well, however, for transactions where the first statement accesses data in the wrong partition or a large number of partitions all at once.

**OP2. Lock only the partitions that the transaction accesses.**

Similarly, knowing all of the partitions that each transaction will access allows the DBMS to avoid traditional concurrency control. If a single-partition transaction will only access data from its base partition, then it can be executed to completion without any concurrency control. Otherwise, the DBMS will "lock" the minimum partitions needed before the transaction starts; partitions that are not involved will process other transactions. Accurately predicting which partitions are needed allows the DBMS to avoid the overhead of deadlock detection and fine-grained row-based locking [18]. But if a transaction accesses an extra partition that was not predicted, then it must be aborted and re-executed. On the other hand, if the DBMS predicts that a transaction will access multiple partitions but only ends up accessing one, then resources are wasted by keeping unused partitions locked.

**OP3. Disable undo logging for non-aborting transactions.**

Since a parallel DBMS replicates state over multiple nodes, persistent logging in these environments is unnecessary [3, 14]. These systems instead employ a transient undo log that is discarded once the transaction has committed [23]. The cost of maintaining this log per transaction is large relative to its overall execution time, especially for those transactions that are unlikely to abort (excluding DBMS failures). Thus, if the DBMS can be guaranteed that a transaction will never abort after performing a write operation, then logging can be disabled for that transaction. This optimization must be carefully enabled, however, since the node must halt if a transaction aborts without undo logging.

This optimization is applicable to all main-memory DBMSs, as undo logging is only needed to abort a transaction and not for recovery as used in disk-based systems. This also assumes that that each procedure's control code is robust and will not abort due to programmer error (e.g., divide by zero).

**OP4. Speculatively commit the transaction at partitions that it no longer needs to access.**

The final optimization that we consider is using speculative execution when a distributed transaction is finished at a partition. For distributed transactions, many DBMSs use two-phase commit to ensure consistency and atomicity. This requires an extra round of network communication: the DBMS sends a prepare message to all partitions and must wait for all of the acknowledgements before it can inform the client that the transaction committed. If the DBMS can identify that a particular query is the last operation that a transaction will perform at a partition, then that query and the prepare message can be combined. This is called the "early prepare" or "unsolicited vote" optimization, and has been shown to improve both latency and throughput in parallel systems [21].

Once a node receives this early prepare for the distributed transaction, the DBMS can begin to process other queued transactions at that node [4, 18]. If these speculatively executed transactions only access tables not modified by the distributed transaction, then they will commit immediately once they are finished. Otherwise, they must wait until the distributed transaction commits. This optimization is similar to releasing locks early in traditional databases' two-phase commit prepare phase [8].

Predicting whether a query is the last one for a given partition is not straightforward for the traditional "conversational" interface because the DBMS does not know what the clients will send next. But even for stored procedures this is not easy, as conditional statements and loops make it non-trivial to determine which queries will be executed by the transaction. As with the other optimizations, the DBMS will have to undo work if it is wrong. If a transaction accesses a partition that it previously declared to be finished with,

```
class NewOrder extends StoredProcedure {
  Query GetWarehouse = "SELECT * FROM WAREHOUSE WHERE W_ID = ?";
  Query CheckStock   = "SELECT S_QTY FROM STOCK
                        WHERE S_W_ID = ? AND S_I_ID = ?";
  Query InsertOrder  = "INSERT INTO ORDERS VALUES (?,?)";
  Query InsertOrdLine = "INSERT INTO ORDER_LINE VALUES (?,?,?,?)";
  Query UpdateStock  = "UPDATE STOCK SET S_QTY = S_QTY - ?
                        WHERE S_W_ID = ? AND S_I_ID = ?";

  int run(int w_id, int i_ids[], int i_w_ids[], int i_qtys[])
    queueSQL(GetWarehouse, w_id);
    for (int i = 0; i < i_ids.length; i++)
      queueSQL(CheckStock, i_w_ids[i], i_ids[i]);
    Result r[] = executeBatch();

    int o_id = r[0].get("W_NEXT_O_ID") + 1;
    queueSQL(InsertOrder, w_id, o_id);
    for (int i = 0; i < r.length; i++) {
      if (r[i+1].get("S_QTY") < i_qtys[i]) abort();
      queueSQL(InsertOrderLine, w_id, o_id, i_ids[i], i_qtys[i]);
      queueSQL(UpdateStock, i_qtys[i], i_w_ids[i], i_ids[i]);
    }
    return (executeBatch() != null);
}
```

**Figure 2:** A stored procedure defines (1) a set of parameterized queries and (2) control code. For each new transaction request, the DBMS invokes the procedure's `run` method and passes in (3) the procedure input parameters sent by the client. The transaction invokes queries by passing their unique handle to the DBMS along with the values of its (4) query input parameters.

then that transaction and all speculatively executed transactions at the partition are aborted and restarted.

### 2.1 Motivating Example

To demonstrate how the above optimizations improve transaction throughput, we consider an example from the TPC-C benchmark [24]. A simplified version of the TPC-C `NewOrder` stored procedure is shown in Fig. 2. Approximately 90% of the `NewOrder` requests create orders using items from a single warehouse. If the database is partitioned by warehouse ids (`w_id`), then most of these requests are executed as single-partitioned transactions [23].

We executed `NewOrder` transactions using H-Store in three different ways: (1) all requests are assumed to be distributed and are executed on a random node locking all partitions; (2) all requests are assumed to be single-partitioned and are executed on a random node, and if the transaction tries to access multiple partitions it is aborted and restarted as a distributed transaction that locks the partitions it tried to access before it was aborted; and (3) the client provides the system with the partitions needed for each request and whether it will abort, and the DBMS only locks the necessary partitions. This last case is the best possible scenario for the DBMS. We execute each configuration using five different cluster sizes, with two partitions/warehouses assigned per node. Transaction requests are submitted from clients executing on separate machines in the cluster. Each trial is executed three times and we report the average throughput of the three runs. in Section 6.

The results in Fig. 3 show the significance of knowing what a transaction will do before it executes in a system like H-Store. The throughput for the "assume distributed" case is constant for all cluster sizes because the DBMS is limited to the rate that it can send and receive the two-phase commit acknowledgements. When there are only a small number of partitions, the other strategies are roughly equivalent because the likelihood that a transaction is on the partition that has the data it needs is higher. The throughput of H-Store, however, scales better when the system has the proper information before a transaction begins, as opposed to restarting a transaction once it deviates from the single-partitioned assumption.

### 3. TRANSACTION MODELS

The throughput improvements in the previous experiment require the application to specify exactly which partitions will be accessed and whether the transaction will abort, which depends on how the data is partitioned and the state of the database. This adds additional burden on developers. Worse, this will change any time the database is reorganized. An alternative approach is to model transactions in such a way that allows the DBMS automatically extract properties for each new transaction and then dynamically enable optimizations without needing to modify the application's code.

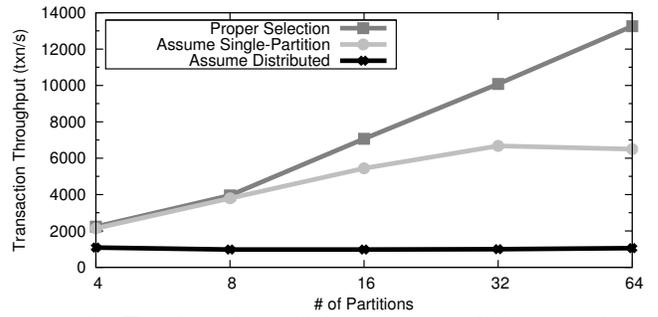

**Figure 3:** The throughput of the system on different partition sizes using three different execution scenarios: (1) All transactions are executed as distributed; (2) All transactions are executed as single-partitioned, distributed transactions are restarted; (3) Single-partition transactions run without concurrency control and distributed transactions lock the minimum number of partitions.

Markov models are an excellent fit for our problem because they can be both generated quickly and used to estimate transaction properties without expensive computations [17]. The latter is important for OLTP systems, since it is not useful to spend 50 ms deciding which optimizations to enable for a 10 ms transaction. In this section, we define our transaction Markov models and outline how they are generated. We describe how to use these models to select optimizations, as well as how to maintain them, in subsequent sections.

### 3.1 Definition

Stored procedures are composed of a set of queries that have unique names. A given invocation of a stored procedure executes a subset of these queries in some order, possibly repeating queries any number of times due to loops. For a stored procedure $\mathcal{SP}_\ell$, we define the transaction Markov model $\mathcal{M}_\ell$ as an acyclic directed graph of the execution states and paths of $\mathcal{SP}_\ell$. An *execution state* is defined as a vertex $v_i \in V(\mathcal{M}_\ell)$ that represents a unique invocation of a single query within $\mathcal{SP}_\ell$, where $v_i$ is identified by (1) the name of the query, (2) the number of times that the query has been executed previously in the transaction (`counter`), (3) the set of partitions that this query will access (`partitions`) as returned by the DBMS's internal API [5], and (4) the set of partitions that the transaction has already accessed (`previous`). In essence, a vertex encodes all of the relevant execution history for a transaction up to that point. Each model also contains three vertices that represent the `begin`, `commit`, and `abort` states of a transaction. Two vertices $v_i, v_j$ are adjacent in $\mathcal{M}_\ell$ through the directed edge $e_{i,j} \in E(\mathcal{M}_\ell)$ if a transaction executes $v_j$'s query immediately after $v_i$'s query.

The outgoing edges from a vertex $v_i \in V(\mathcal{M}_\ell)$ represent the probability distribution that a transaction transitions from $v_i$'s state to one of its subsequent states. If a transaction committed, then the vertex for the last query it executed is connected by an edge to the `commit` state; in the same way, if the transaction aborted, then the last query's vertex is connected to the `abort` state. A transaction's *execution path* in $\mathcal{M}_\ell$ is an ordered list of vertices from the `begin` state to one of these two terminal states.

These Markov models are used to predict the future states of new transactions based on the history of previous transactions. Each model is generated from a sample *workload trace* for an application. A trace contains for each transaction (1) its procedure input

87

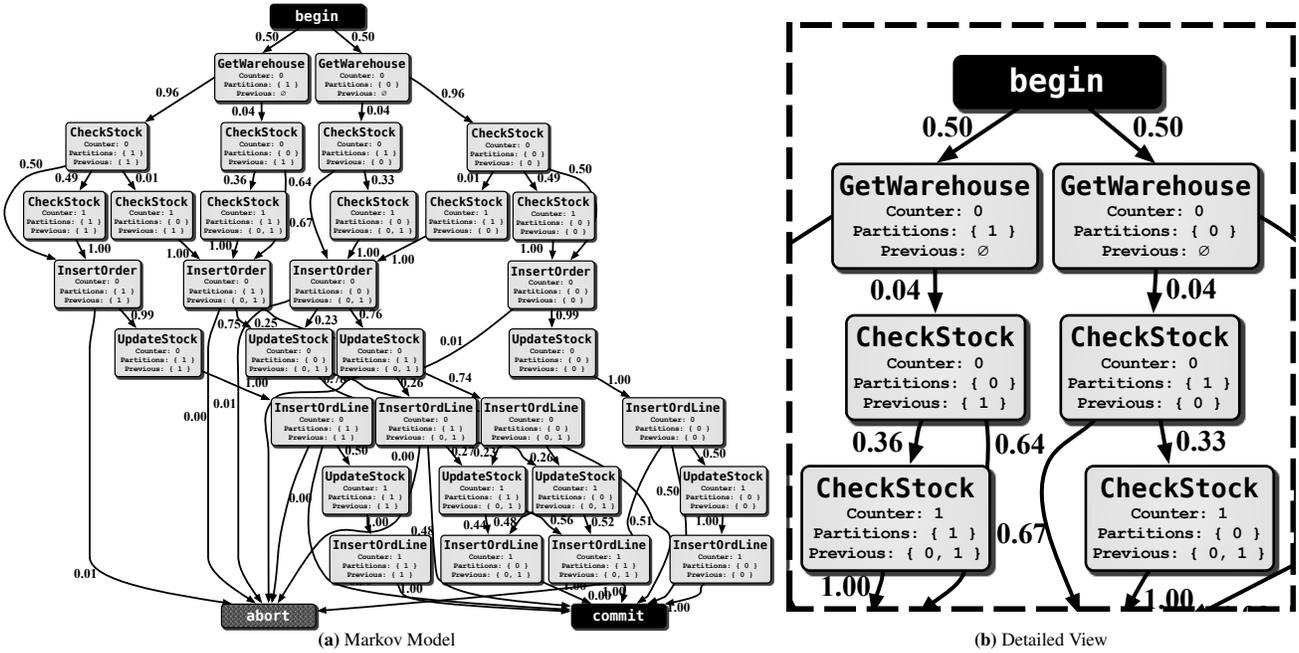

**Figure 4:** An example of a Markov model for the `NewOrder` stored procedure shown in Fig. 2.

parameters and (2) the queries it executed, with their corresponding parameters. Because the trace does not encode what partitions each query accessed, new models must be regenerated from the trace whenever the database's partitioning scheme changes.

Fig. 4 shows an example of a Markov model for the `NewOrder` procedure in Fig. 2. In the detailed view shown in Fig. 4b, we see that there are two `GetWarehouse` vertices that are adjacent to the `begin` vertex. The sets of previously accessed partitions for these vertices are empty since they are the first query in the transaction, while the subsequent `CheckStock` states include partitions that were touched by their parent vertices. For simplicity, Fig. 4 was generated for a database that has only two partitions, and thus every `NewOrder` transaction executes the `GetWarehouse` query on just one of the two partitions (assuming that each warehouse is assigned to one partition).

Every vertex is also annotated with a table of probabilities for events that may occur after the transaction reaches that particular state. This is used to make initial predictions about a transaction, and to refine and validate those predictions as it executes. The table's values are derived from the probability distributions of the state transitions inherent in a Markov model, but are pre-computed in order to avoid having to perform an expensive traversal of the model for each transaction. This step is optional but reduces the on-line computing time for each transaction by an average of 24%, which is important for short-lived transactions. As shown in Fig. 5, a probability table contains two types of estimates. The first type are global predictions on (1) the probability that the transaction's future queries will execute on the same partition as where its control code is executing (**OP1**) and (2) the probability that the transaction will abort (**OP3**). For each partition in the cluster, the table also includes the probability that a transaction will execute a query that either reads or writes data at that partition (**OP2**), or conversely whether a transaction is finished at that partition (**OP4**).

### 3.2 Model Generation

A stored procedure's Markov model is generated in two parts. In the first part, called the *construction phase*, we create all known execution states from the workload trace. Next, in the *processing phase*, we traverse the model and calculate its probability distribu-tions. We discuss adding new states at run time in Section 4.4.

**Figure 5:** The probability table for the `GetWarehouse` state from Fig. 4. The table shows that with 100% certainty any transaction that reaches this state will execute another query that accesses partition #0 before it commits. Conversely, there is a 5% chance that it will need to either read or write data on partition #1.

**Construction Phase:** A new model for a transaction initially contains no edges and the three vertices for the `begin`, `commit`, and `abort` states. For each transaction record in the workload trace, we estimate the partitions accessed by its queries using the DBMS's internal API for the target cluster configuration [5]. We then traverse the corresponding path in the model, adding vertices and edges where appropriate. After all queries in the transaction have been processed, the last vertex in the transaction's path is connected to one of the terminal states. At the end of this phase, all of the initial execution states and edges have been created.

**Processing Phase:** In terms of the model, an edge's probability represents the likelihood that a transaction at the parent vertex will transition along the edge to the child vertex. In terms of the transaction, this is the probability that a transaction that has reached the parent state will execute the child's query next. The processing phase visits each vertex in the model, assigning probabilities to each outgoing edge. The probability is computed as the number of times an edge was visited divided by the total number of times the vertex was reached in the construction phase.

After the edge probabilities are calculated, we then pre-compute the vertex probability tables. A vertex's probability table is based on its children's tables weighted by their edge probabilities. The first step is, therefore, to initialize the default probabilities in the terminal states: all of the partition-specific probabilities at the `commit` vertex and the global abort probability at the `abort` vertex are both set to one. Then to calculate the tables for the remaining vertices,

88

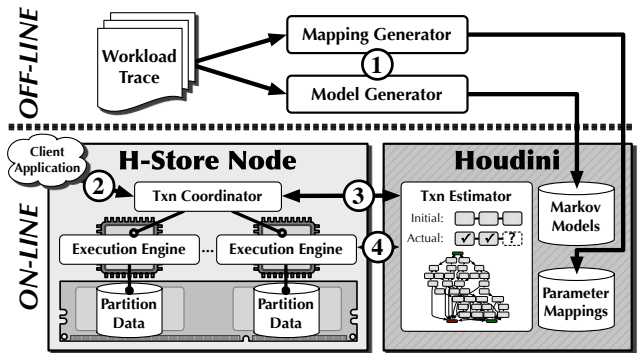
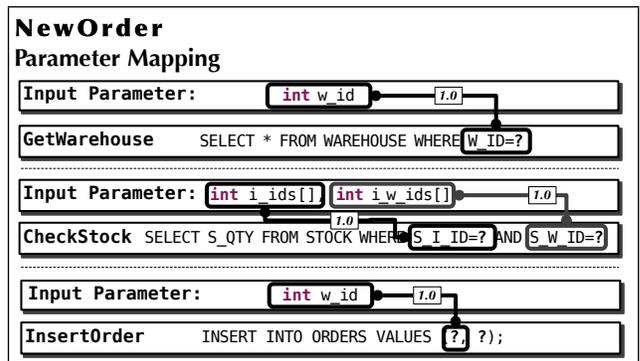

**Figure 6:** An overview of the Houdini predictive framework: (1) at initialization time, Houdini generates the Markov models and parameter mappings using a workload trace; (2) at run time, the client sends transaction requests to the DBMS's transaction coordinator; (3) the DBMS passes this request to Houdini, which generates an initial path estimate and selects optimizations; (4) Houdini monitors the transaction as it executes and provides updates to the DBMS.

**Figure 7:** A parameter mapping for the `NewOrder` procedure.

we traverse the model in ascending order based on the length of the longest path from each vertex to either the `commit` or `abort` vertex. Traversing the graph in this manner ensures that a vertex's table is only calculated after the tables for all of its children have been calculated. If the query at a vertex reads or writes data at a particular partition, then the corresponding entry in that vertex's probability table for that partition is set to one and the finish probability is set to zero. For those partitions not accessed at a state, then the read/write/finish probabilities are the sum of their children vertices' table entries at that partition weighted on the edge probabilities to each of those child vertices.

## 4. PREDICTIVE FRAMEWORK

Given this definition of our Markov models, we now present *Houdini*, a framework for "magically" predicting the actions of transactions at run time. Such a framework can be embedded in a DBMS to enable it to automatically optimize its workload. Houdini's functionalities are designed to be autonomous, and thus do not require human intervention to maintain once it is deployed.

As shown in Fig. 6, Houdini is deployed on each node in the cluster and is provided with all of the Markov models generated off-line for the application's stored procedures. When a transaction request arrives at a node, the DBMS passes the request (i.e., procedure name and input parameters) to Houdini, which then generates an *initial estimate* of the transaction's execution path. This path represents the execution states that the transaction will likely reach in the Markov model for that procedure. From this initial path, Houdini informs the DBMS which of the optimizations described in Section 2 to enable for that request.

Determining the initial properties of a transaction before it executes is the critical component of our work. We first describe a technique for mapping the procedure input parameters to query parameters so that we can predict what partitions queries will access. We then describe how to construct the initial path in our Markov models using these parameter mappings and how Houdini uses it to select which optimizations to enable. Lastly, we discuss how Houdini checks whether the initial path matches what the transaction does and makes adjustments in the DBMS.

### 4.1 Parameter Mappings

We first observe that for most transactions in OLTP workloads, the set of partitions that each query will access is dependent on its input parameters and the database's current state [23]. A corollary to this is that the query parameters that are used in predicates on tables' partitioning attributes are often provided as procedure input parameters, and therefore they are not dependent on the output of earlier queries in the transaction. For example, the first input parameter to Fig. 2 is the warehouse id (`w_id`) that is used as an input parameter for almost all of the queries in `NewOrder`. Given this, for those queries whose input parameters that are "linked" to procedure parameters, we can determine what partitions the queries will access using the values of the procedure parameters at run time. Although procedures that do not follow this rule do exist, in our experience they are the exception in OLTP applications or are the byproduct of poor application design.

To capture such relationships, we use a data structure called a *parameter mapping* that is derived from the sample workload trace. A procedure's parameter mapping identifies (1) the procedure input parameters that are also used as query input parameters and (2) the input parameters for one query that are also used as the input parameters for other queries. We use a dynamic analysis technique to derive mappings from a sample workload trace. One could also use static analysis techniques, such as symbolic evaluation or taint checking, but these approaches would still need to be combined with traces using dataflow analysis since a transaction's execution path could be dependent on the state of the database.

To create a new mapping for a procedure, we examine each transaction record for that procedure in the workload and compare its procedure input parameters with all of the input parameters for each query executed in that transaction. For each unique pairwise combination of procedure parameters and query parameters, we count the number of times that the two parameters had the same value in a transaction. After processing all of the records in this manner, we then calculate the *mapping coefficient* for all parameter pairs as the number of times that the values for that pair were the same divided by the number of comparisons performed. As shown in the example in Fig. 7, the first procedure parameter has the same value as the first query parameter for `GetWarehouse` (i.e., the mapping coefficient is equal to one), and thus we infer that they are the same variable in the procedure's control code. We apply this same technique to the other queries and map their input parameters as well.

A parameter mapping also supports transactions where the same query is executed multiple times and when the stored procedure has non-scalar input parameters. If a query is executed multiple times in the same transaction, then each invocation is considered a unique query. Likewise, if a procedure input parameter is an array, then each element of that array is treated as a unique parameter. From the mapping in Fig. 7, we identify that the $n$-th element of the `i_ids` array is linked to the third parameter of the $n$-th invocation of `InsertOrdLine` in Fig. 2. For each element in a procedure parameter array, we compare it with all of the query parameters within the current transaction just as before. The coefficients for

89

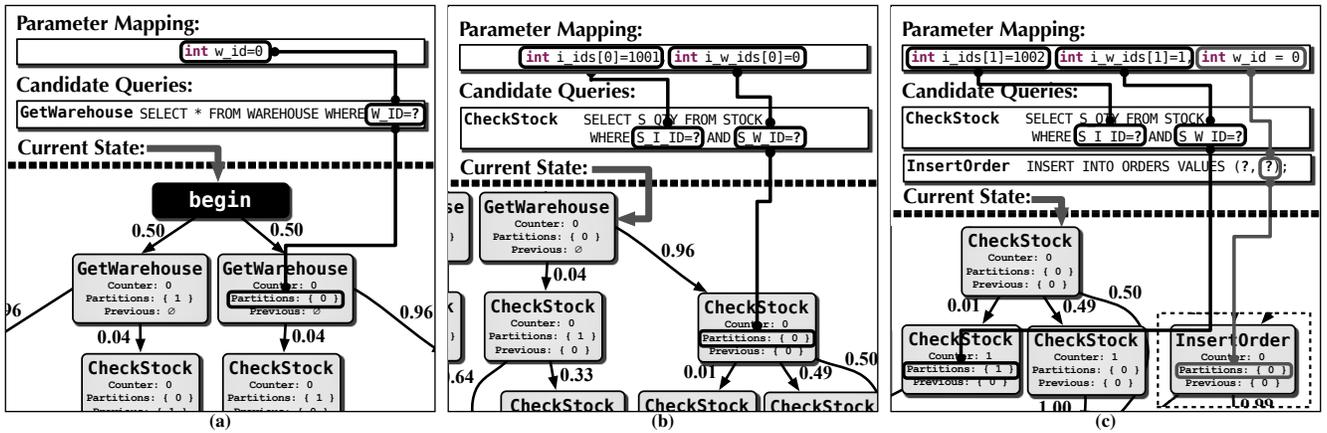

**Figure 8:** An example of generating the initial execution path estimate for a `NewOrder` invocation. As shown in the trace record in Fig. 7, the procedure parameters in this example are (**w_id**=0, **i_ids**=[1001,1002], **w_i_ids**=[0,1], **i_qtys**=[2,7]).

multiple query instances or array parameters are aggregated into a single value using their geometric mean.

We remove false positives by discarding any mapping coefficients that are below a threshold; these occur when parameters randomly have the same values or when the control code contains a conditional block that modifies the input parameter. We found empirically that coefficients greater than 0.9 seem to all give the same result for the workloads that we investigated.

## 4.2 Initial Execution Path Estimation

Now with the procedure parameter mappings, Houdini constructs the initial execution path estimate in the Markov models for each new transaction request that arrives at the DBMS.

To generate a path estimate for a transaction, we first enumerate all of the successor states to the `begin` state and construct the set of candidate queries. We then estimate which partitions these candidates queries will access using the procedure's parameter mapping. This determines whether transitioning from the current state to the state represented by these queries (and the set of partitions that they access) is valid. A state transition is *valid* for a transaction if (1) we can determine all the query parameters needed for calculating the partitions accessed by that state's query and (2) the next state's set of previously accessed partitions contains all the partitions accessed by the transaction up to this point. For those transitions that are valid, we choose the one with the greatest edge probability, append this state to the initial path estimate, and then repeat the process.

We now illustrate these steps using the `NewOrder` Markov model shown in Fig. 4. As shown in Fig. 8a, the only candidate query when a transaction starts is `GetWarehouse`. Using the parameter mapping shown in Fig. 7, we identify that `GetWarehouse`'s first parameter is mapped to the procedure's first parameter (`w_id`). Therefore, we can compute which partitions `GetWarehouse` accesses because we know with a high-degree of certainty the value of its only input parameter is the same as the value of `w_id`. We then select the next state as the particular `GetWarehouse` state that accesses the same partitions as was estimated. This process is repeated in the next step in Fig. 8b: the candidate query set contains only `CheckStock`, so we again use the mapping to get the values of the procedure parameters that are used for that query and compute which partitions that the query accesses. We then select the next state in the path as the one that represents the first invocation of `CheckStock` that accesses these partitions and also has the correct previously accessed partitions for the transaction.

The technique described above works well when either the procedure's control code is linear or all the pertinent query parameters are mappable to procedure parameters, whereupon the Markov model is essentially a state machine. But many procedures contain conditional branches, and thus it is not always possible to resolve which state is next simply by estimating partitions. An example of this is shown in Fig. 8c. There are two choices for which query that the transaction could execute next: (1) the second invocation of `CheckStock` or (2) the first invocation of `InsertOrder`. Both transitions are valid if the size of the `i_ids` procedure input parameter array is greater than one. If the size of this array was one, then Houdini would infer that the transaction could never execute `CheckStock` a second time. When such uncertainty arises, we chose the edge with the greater weight.

The path estimate is complete once the transaction transitions to either the `commit` or `abort` state.

## 4.3 Initial Optimizations Selection

Using a transaction's initial path estimate, Houdini chooses which optimizations the DBMS should enable when it executes the transaction. We now describe how Houdini selects these optimizations.

For each potential optimization, we calculate a *confidence coefficient* that denotes how likely that it is correct. This coefficient is based on the probabilities of the edges selected in the transaction's initial path estimate. Houdini prunes estimations if their corresponding confidence is less than a certain threshold. Setting this threshold too high creates false negatives, preventing the DBMS from enabling valid optimizations. Conversely, setting this threshold too low creates false positives, causing the DBMS to enable certain optimizations for transactions that turn out to be incorrect and therefore it will have to rollback work. We explore the sensitivity of this threshold in our evaluation in Section 6.5.

**OP1.** Houdini counts each time that a partition is accessed by a query in the transaction's initial path estimate. The partition that is accessed the most is selected as the transaction's base partition.

**OP2.** Similarly, the set of partitions that the transaction needs (and therefore the DBMS should lock) is the based on the execution states in the initial path estimate. The probability that a partition is accessed is the confidence coefficient of the edges in the initial path up to the first vertex that accesses that partition.

**OP3.** Because multi-partition and speculatively executed transactions can be aborted as a result of other transactions in the system, these transactions are always executed with undo logging enabled. Thus, Houdini will only determine which non-speculative single-partition transactions can be executed without undo buffers. Houdini is more cautious when estimating whether transactions could abort because unlike the other optimizations, it will be very expensive to recover if it is wrong. To avoid this, we use the greatest abort probability in the all of the tables in the initial path estimate. That is, the probability that the transaction will abort is the largest abort probability value in all of the states' tables.

### 4.4 Optimization Updates

After creating the initial path and optimization estimates for a transaction, Houdini provides this information to the DBMS. The transaction is then queued for execution at the current node or redirected based on the estimate. Once the transaction starts, Houdini tracks its execution and constructs the path of execution states that the transaction enters in its stored procedure's model. At each state, Houdini (1) determines whether the transaction has deviated from the initial path estimate and (2) derives new information based on the transaction's current state. If the transaction reaches a state that does not exist in the model, then a new vertex is added as a placeholder; no further information can be derived about that state until Houdini recomputes the model's probabilities (Section 4.5). Otherwise, Houdini uses the current state to provide updates to the DBMS's transaction coordinator:

**OP3**. Houdini uses the pre-calculated probability tables to check whether a single-partition transaction has reached a point in its control code that will never abort (i.e., there is no path from the current state to the `abort` state). When this occurs, the DBMS disables undo logging for the remainder of the transaction's execution.

**OP4**. Houdini also uses the probability tables to determine whether a distributed transaction is finished with partitions. If the finish probability for a particular partition is above the confidence threshold, then Houdini informs the DBMS that the transaction no longer needs that partition. This allows the DBMS to send the early prepare message [21] and speculatively execute transactions at these partitions [4, 18]. If the transaction was read-only at a partition, then it commits immediately and the DBMS begins to execute other transactions on that partition. Otherwise, the speculative transaction waits until the distributed transaction finishes.

### 4.5 Model Maintenance

The probability that a transaction transitions from one state to another is based on static properties of the sample workload trace that was used to generate the models. If an application's workload shifts, then the models may no longer represent the current behavioral state of that application. For example, if previous `New-Order` transactions in the trace only inserted two items but now incoming requests have three or more, then Houdini will incorrectly choose initial paths that only executed the `CheckStock` query twice. Houdini can identify when the workload has changed [15] and to adjust to these changes without having to re-create the models. This occurs on-line without stopping the system; new models only need to be generated off-line when the database's partitioning scheme changes or when the procedure's control code is modified.

Houdini determines whether a model is no longer accurate by measuring how often it chooses a state transition for transactions that does not match the expected edge probability distribution. As a transaction executes, Houdini constructs its actual execution path in the model and increments internal counters whenever the transaction "visits" an edge. As long as the distribution of the transitions from each vertex is within some threshold of the original probabilities in the Markov model, then Houdini infers that the model is still in sync with the application. If the distribution no longer matches the model's expectations, then Houdini recalculates the edge and vertex probabilities based on the edge counters. Since this is an inexpensive operation ($\leq 5$ ms), our current implementation uses a threshold of 75% accuracy before Houdini recomputes the probabilities. We defer the exploration of more robust techniques as future work, such as a sliding window that only includes recent transactions for fast changing workloads.

### 4.6 Limitations

There are three ways that Houdini may fail to improve the throughput of the DBMS. The first case is if the overhead from calculating the initial path estimate negates the optimizations' performance gains. This can occur if a model is very wide (i.e., many transition possibilities per state) or very long (i.e., many queries executed per transaction). For the latter, the limit is approximately 175-200 queries per transaction in our current implementation. It simply takes too long for Houdini to traverse the model for these transactions and compute the partitions that could be accessed at each state. Pre-computing initial path estimates for stored procedures that are always single-partition would alleviate this problem to some extent, but it is not applicable for procedures that are distributed only some of the time since Houdini needs the path estimate to determine what partitions will be accessed. We note, however, that procedures that execute many queries and touch a large portion of the database are not the main focus of high-performance OLTP systems.

Additionally, storing all of the execution states for a stored procedure in a single "global" Markov model can be difficult to scale for large clusters. The total number of states per model is combinatorial for procedures like `NewOrder` that access combinations of partitions, most of which are unreachable based on where the transaction's control code is executing. For example, a transaction executing at a particular partition can only reach just one of the `Get-Warehouse` states in Fig. 4b and based on which one that is, other states can never be reached. These global models are also problematic on multi-core nodes, since Houdini must either use separate copies of the models for each execution thread, or use locks to avoid consistency issues when it updates the models.

The last type of limitation that can hinder DBMS performance is if the models are unable to accurately predict what a transaction will do, causing the DBMS to make wrong decisions and possibly have to redo work. As an example of this, consider a `NewOrder` request that has two items to insert from different warehouses (i.e., partitions). If Houdini uses the Markov model in Fig. 4 to predict this transaction's initial path, then it would not select the correct execution state from the choices shown in Fig. 8c. This is because the second invocation of `CheckStock` and the `InsertOrder` query are both valid states; the length of the length of the warehouse id array (`i_w_ids`) is greater than one, and thus the transaction could potentially execute either query. As described in Section 4.2, when such uncertainty arises, we choose the next state transition based on edge with the greatest probability. This is still insufficient, however, since the probability of the transition that the transaction will actually take is less than the other potential transition. The Markov model in Fig. 4 does not capture the fact that the number of `CheckStock` queries corresponds to the length of the `i_w_ids` array. This is problematic in our example because the query that Houdini failed to predict in the model accesses a partition that is different than the ones from the transaction's previous queries.

## 5. MODEL PARTITIONING

Given these limitations, we now describe how Houdini automatically partitions the Markov models for a given application to improve their prediction efficacy and scalability. Houdini clusters the transactions for each procedure in the sample workload trace based on salient attributes of its input parameters. This allows us to cap-

91

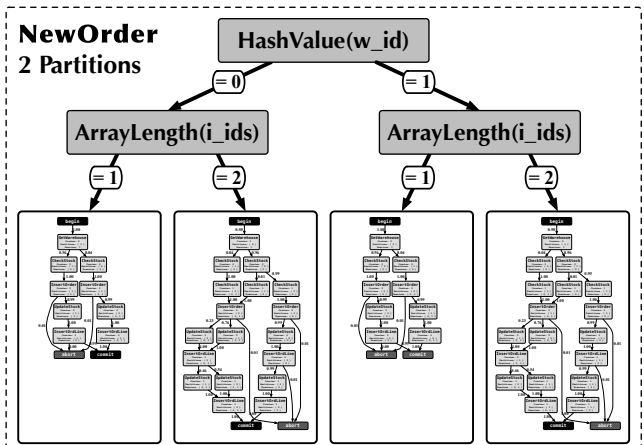

**Figure 9:** A partitioned set of `NewOrder` Markov models. The decision tree above the models divides transactions by the hash value of the first procedure parameter and the length of the array of the second procedure parameter. The detail of the models in the above figure is not relevant other than to note that they are less complex than the global model for the same procedure shown in Fig. 4.

| Feature Category | Description |
| --- | --- |
| NORMALIZEDVALUE($x$) | The normalized value of the parameter $x$. |
| HASHVALUE($x$) | The hash value of the parameter $x$. |
| ISNULL($x$) | Whether the value of the parameter $x$ is null. |
| ARRAYLENGTH($x$) | The length of the array parameter $x$. |
| ARRAYALLSAMEHASH($x$) | Whether all elements of the array parameter $x$ hash to the value. |

**Table 1:** The list of feature categories that are extracted from the stored procedure input parameters for each transaction trace record. These features are used when sub-dividing the models for each stored procedure to improve scalability and prediction accuracy.

| Feature Instance | Value | Feature Instance | Value |
| --- | --- | --- | --- |
| HASHVALUE(w_id) | 0 | ARRAYLENGTH(w_id) | null |
| HASHVALUE(i_ids) | null | ARRAYLENGTH(i_ids) | 2 |
| HASHVALUE(i_w_id) | null | ARRAYLENGTH(i_w_ids) | 2 |
| HASHVALUE(i_qtys) | null | ARRAYLENGTH(i_qtys) | 2 |

**Table 2:** The feature vector extracted from the transaction example in Fig. 8. The value for the ARRAYLENGTH(w_id) feature is null because the w_id procedure parameter in Fig. 2 is not an array.

ture certain nuances of the transactions, such as variability in the size of input parameter arrays. As shown in Fig. 9, we generate models for each of these clusters and support them with a decision tree that allows Houdini to quickly select the right model to use for each incoming transaction request at run time.

Dividing the models in the manner that we now describe is a well-known and effective technique from the machine learning and optimization communities [26, 19].

### 5.1 Clustering

The goal of the clustering process is to group transactions together based on their features in such a way that the Markov models for each cluster more accurately represent the transactions. We define a *feature* in this context as an attribute that is derived from a transaction's stored procedure input parameters [12]. For example, one feature could be the length of the array for one particular parameter, while another could be whether all of the values in that array are the same. Table 1 shows the different categories of features that are extracted from transaction records. A *feature vector* is a list of values for these features that are extracted from each transaction trace record in the sample workload. Each transaction's feature vector contains one value per input parameter per category.

An example of a feature vector is shown in Table 2

After extracting the feature vectors for each of the transaction records in the workload trace, we then employ a machine learning toolkit to cluster the transactions of each procedure based on these vectors [13]. We use the *expected maximization* clustering algorithm, as it does not require one to specify the number of clusters beforehand. The transaction records are each assigned to a cluster by this algorithm and then we train a new Markov model that is specific for each cluster using these records. For example, if we clustered the `NewOrder` transactions based on the length of the i_w_ids input parameter, then the number of `CheckStock` invocations for all transactions in each cluster will be the same.

### 5.2 Feed-Forward Selection

The problem with the above clustering approach is that it is decoupled from Houdini's ability to predict a transaction's properties accurately using the models; that is, the clustering algorithm may choose clusters for a stored procedure based on features that do not improve the accuracy of the models' predictions compared to the single non-clustered model. Therefore, we need to determine which set of features are relevant for each procedure in order to cluster the transactions properly. Enumerating the power set of features with a brute-force search to evaluate the accuracy of all feature combinations is not feasible, since the amount of time needed to find the optimal feature set is exponential. This would simply take too long for applications either with a large number of stored procedures or with stored procedures that have many input parameters.

We instead use a greedy algorithm called *feed-forward selection* as a faster alternative [12, 19]. This algorithm first iterates all unique feature combinations for small set sizes and then constructs larger sets using only those features that were in the smaller sets that improve the predictions. In each round $r$, we create all sets of features of size $r$ and measure how well they predict the initial execution paths of transactions. After each round, we sort the feature sets in ascending order and select the features in the top 10% sets with the best accuracy. We repeat the process in the next round using sets of size $r+1$. The search stops when at the end of a round the algorithm fails to find at least one feature set that produces clustered models with better prediction accuracy than the best feature set found in the previous rounds.

To begin, we first split the sample workload for the target stored procedure into three disjoint segments, called the *training workset* (30%), the *validation workset* (30%), and the *testing workset* (40%) [19]. Then we enumerate the power set of features for the current round (e.g., if there $n$ features, then the initial round will have $n$ one-element sets). For each feature set in the round, we seed the clustering algorithm on that set using the training workset. We then use the seeded clusterer to divide the transaction records in the validation workset and generate the Markov models for each cluster using the same method described in Section 3.2.

Now with a separate Markov model per cluster for a particular feature set, we estimate the accuracy of the clustered models using the remaining records in the testing workset. For each of these transaction records, we generate an initial path estimate using Houdini just as if it was a new transaction request and then simulate the transaction executing in the system by generating the "actual" execution path of the transaction. We measure the accuracy of these initial path estimates not only based on whether it has the same execution states as the actual path, but also based on whether Houdini correctly generates transaction updates.

The accuracy for each initial path estimate is based on the optimizations defined in Section 2. The penalty for incorrectly predicting that a single-partition transaction will not abort is infinite, since

92

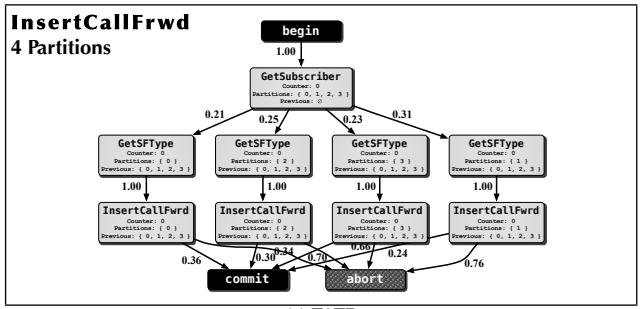

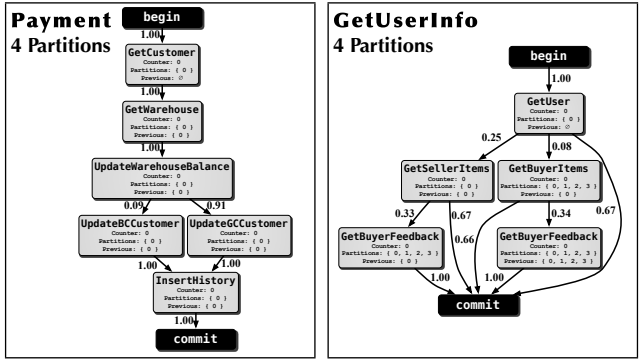

(a) TATP  (b) TPC-C  (c) AuctionMark

**Figure 10:** Markov models for select stored procedures from the three OLTP benchmarks used in our evaluation.

it puts the database in an unrecoverable state. The total accuracy measurement for each feature set is the sum of these penalties for all transactions in the testing workset.

### 5.3 Run Time Decision Tree

After the search terminates, we use the feature set with the lowest cost (i.e., most accurately models the transactions for the target stored procedure) to generate a decision tree for the models using the C4.5 classifier algorithm from the same machine learning toolkit [13]. When a new transaction request arrives at the DBMS at run time, Houdini extracts the feature vector for the transaction and traverses this decision tree to select which Markov model to use for that request. For example, the Markov models shown Fig. 9 for the NewOrder stored procedure are clustered on the value of the w_id parameter and the length of the i_w_ids array. The models in the leaf nodes of the tree are specific to these features. This mitigates the scaling, concurrency, and accuracy problems from using a single model per procedure.

### 6. EXPERIMENTAL EVALUATION

We have integrated our modeling algorithms and prediction framework in the H-Store system and now present an evaluation of its usefulness. We use three OLTP benchmarks that have differing levels of complexity in their workloads. We assume that the databases for each benchmark are partitioned in such way that it maximizes the number of single-partition transactions [7]. For each benchmark, we generate sample workload traces of 100,000 transactions collected over a simulated one hour period.

All of the experiments measuring throughput were conducted on a cluster at the University of Wisconsin-Madison. Each node has a single 2.4GHz Intel Core 2 Duo processor with 4GB RAM.

### 6.1 Benchmarks

**TATP:** The Telecom Application Transaction Processing benchmark is an OLTP testing application that simulates a caller location system used by telecommunication providers [25]. TATP has seven stored procedures, of which four are always single-partitioned. The other three each first execute a broadcast query that finds a unique value from a column that the tables are not partitioned on, and then perform some operation at a single partition using that value.

|  |  | TATP | TPC-C | AuctionMark |
|---|---|---|---|---|
| **OP1** | Global | 95.0% | 94.8% | 94.9% |
|  | Partitioned | 94.9% | 99.9% | 94.7% |
| **OP2** | Global | 98.9% | 90.9% | 90.7% |
|  | Partitioned | 100% | 99.0% | 95.4% |
| **OP3** | Global | 100% | 100% | 100% |
|  | Partitioned | 100% | 100% | 100% |
| **OP4** | Global | 99.5% | 100% | 100% |
|  | Partitioned | 99.5% | 95.8% | 99.9% |
| **Total** | Global | **94.9%** | **93.8%** | **85.6%** |
|  | Partitioned | **94.9%** | **95.0%** | **90.2%** |

**Table 3:** Measurements of the global and partitioned Markov models' accuracy in predicting the execution properties of transactions.

**TPC-C:** This benchmark is the current industry standard for evaluating the performance of OLTP systems [24]. It consists of five stored procedures that simulate a warehouse-centric order processing application. The key aspect about this benchmark is that the two most executed procedures vary in whether their transactions touch multiple partitions or not.

**AuctionMark:** AuctionMark is an OLTP benchmark being developed by Brown University and a well-known Internet auction company [1]. It consists of 10 stored procedures, two of which are periodically executed to process recently ended auctions. Many of the procedures involve an interaction between a buyer and a seller whose data are likely stored at different partitions. Other transactions contain conditional branches that execute different queries based on the procedure's input parameters.

### 6.2 Model Accuracy

We first calculated the off-line accuracy of the optimization estimates generated by Houdini for a simulated cluster of 16 partitions. The accuracy of an estimate is based on whether Houdini (1) identifies the optimizations at the correct moment in the transaction's execution (e.g., disabling undo logging at the right time – **OP3**), (2) does not cause the DBMS to perform unnecessary work (e.g., locking partitions that are never used – **OP1**, **OP2**), and (3) does not cause the transaction to be aborted and restarted (e.g., accessing a partition after it was deemed finished – **OP4**). For each procedure, we generate a single "global" model and a set of "partitioned" models using the first 50,000 transaction records from the sample workloads. We then use Houdini to estimate optimizations for the remaining 50,000 transactions. We reset the models after each estimation so as to not learn about new execution states, which would mask any deficiencies.

The results in Table 3 show that the global Markov models enable accurate path estimates for 91.0% of the transactions evaluated, while the partitioned models improved the accuracy rate to 93.4%. Although Houdini fails to select the base partition (**OP1**) for 5% of TATP's transactions, they are all distributed transactions that either update every partition (i.e., the base partition does not matter) or access a single partition based on the result of a multi-partition query (i.e., the best base partition depends on the state of the database). The accuracy for TPC-C is nearly perfect in the partitioned models for **OP1-3**, but that it can miss that a transaction is with finished a partition (**OP4**). The accuracy for AuctionMark transactions is also high, except for the two procedures with conditional branches. Houdini never mispredicts that a transaction will not abort for any benchmark, but it does miss a small number of transactions ($<$1%) where it could have disabled undo logging (**OP3**).

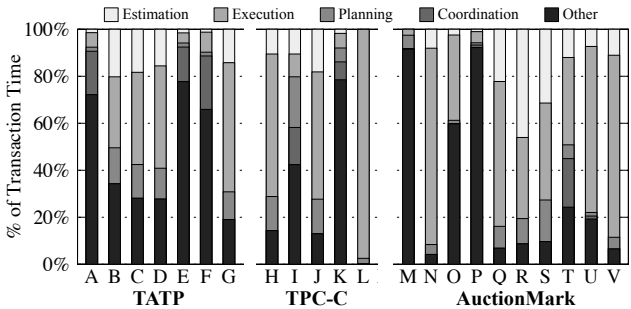

**Figure 11:** Relative measurements of the time spent for each transaction (1) estimating optimizations, (2) executing, (3) planning, (4) coordinating its execution, and (5) other setup operations.

| | Procedure | OP1 | OP2 | OP3 | OP4 | Estimate |
|---|---|---|---|---|---|---|
| **TATP** A | DeleteCallFwrd | - | 100% | - | - | 0.02 ms |
| B | GetAccessData | 98.5% | 100% | 64.8% | 33.7% | 0.01 ms |
| C | GetNewDest | 100% | 100% | 66.4% | 33.6% | 0.01 ms |
| D | GetSubscriber | 98.9% | 100% | 64.9% | 34.1% | 0.01 ms |
| E | InsertCallFwrd | - | 100% | - | - | 0.04 ms |
| F | UpdateLocation | - | 100% | - | - | 0.01 ms |
| G | UpdateSubscriber | 100% | 100% | - | 53.2% | 0.02 ms |
| **TPC-C** H | Delivery | 100% | 100% | 78.6% | 22.4% | 4.23 ms |
| I | NewOrder | 99.5% | 93.2% | 72.5% | 19.6% | 0.43 ms |
| J | OrderStatus | 100% | 100% | 89.6% | 85.3% | 0.05 ms |
| K | Payment | 99.1% | 99.7% | 60.6% | 16.4% | 0.08 ms |
| L | StockLevel | 99.2% | 100% | 46.7% | 22.0% | 0.05 ms |
| **AuctionMark** M | CheckWinningBids | - | - | - | - | - |
| N | GetItem | 100% | 100% | 89.0% | 11.0% | 0.04 ms |
| O | GetUserInfo | 99.9% | 100% | 75.3% | 8.4% | 0.05 ms |
| P | GetWatchedItems | 100% | 100% | - | - | 0.04 ms |
| Q | NewBid | 100% | 100% | 83.2% | 13.3% | 0.26 ms |
| R | NewComment | 99.5% | 100% | 44.6% | 11.3% | 0.13 ms |
| S | NewItem | 100% | 100% | 95.9% | 4.1% | 0.20 ms |
| T | NewPurchase | 99.0% | 100% | 46.4% | 11.1% | 0.12 ms |
| U | PostAuction | - | 55.0% | - | 16.7% | 0.32 ms |
| V | UpdateItem | 100% | 100% | 90.5% | 9.5% | 0.04 ms |

**Table 4:** The percentage of transactions that Houdini successfully enabled one of the four optimizations. In the case of **OP4**, the measurement represents how many transactions were speculatively executed as a result of the early prepare optimization. The rightmost column contains the average amount of time that Houdini spent calculating the initial optimization estimates and updates at run time.

### 6.3 Estimation Overhead

Next, we measured the overhead of using Houdini to estimate the optimizations at run time. We implemented a profiler [14] that records the amount of time H-Store spends for each transaction (1) estimating the initial execution path and updates, (2) executing its control code and queries, (3) planning its execution, (4) coordinating its execution, and (5) miscellaneous setup operations. We executed the benchmarks on a 16-partition H-Store cluster and report the average time for each of these measurements. Profiling begins when a request arrives at a node and then stops when the result is sent back to the client. We use the partitioned models so that the cost of traversing the decision tree is included in the measurements.

The results in Fig. 11 show that only an average of 5.8% of the transactions' total execution time is spent in Houdini. This time is shared equally between estimating the initial path versus calculating updates. All procedures with an overhead greater than 15% are short-lived single-partitioned transactions. For example, 46.5% of AuctionMark NewComment's execution time is spent selecting optimizations, but it is the shortest transaction (i.e., average execution time is just 0.29 ms). Although we do not discuss such techniques in this paper, Houdini can completely avoid this if it caches the estimations for any non-abortable, always single-partition transactions.

### 6.4 Transaction Throughput

We next measured the throughput of H-Store when deployed with Houdini. We execute each benchmark using five different cluster sizes, with two partitions assigned per node. Transaction requests are submitted from multiple client processes running on separate machines in the cluster. We use four client threads per partition to ensure that the workload queues at each node are always full. We execute each benchmark three times per cluster size and report the average throughput of these trials. In each trial, the DBMS is allowed to "warm-up" for 60 seconds and then the throughput is measured after five minutes. As H-Store executes, we record the percentage of transactions for each procedure where Houdini successfully selected an optimization. Note that this is different than the accuracy measurements shown Table 3, because Houdini now must consider the run time state of the DBMS (e.g., it cannot disable undo logging for speculative transactions).

We executed the benchmarks with Houdini first using the global Markov models and then again using the partitioned models. We allow Houdini to "learn" about new execution states in the models in the warm-up period, and then recompute the probabilities before running the measured workload. For each new transaction request, Houdini generates the initial path estimate and determines whether the request needs to be redirected to a different node (**OP1**) and can be executed with undo logging (**OP3**). As the transaction executes, Houdini checks whether it is finished with other partitions (**OP4**) or no longer needs undo logging (**OP3**). Any transaction that attempts to access a partition that Houdini failed to predict (**OP2**) is aborted and restarted as a multi-partition transaction that locks all partitions.

To compare how H-Store performs without Houdini, we also executed the benchmarks using DB2-style transaction redirects [6]. When operating in this mode, the DBMS first executes every request as single-partition transaction at a random partition on the node where the request originally arrived. If a transaction attempts to access a partition that is different than the one it was assigned to, then it is aborted and redirected to the correct node. If the transaction attempts to access multiple partitions, none of which are at the node where it is currently executing at, then it is redirected to the partition that it requested the most and is executed as a multi-partition transaction (with random tiebreakers). Because the DBMS has no way to infer the transaction properties without Houdini, it cannot use the other optimizations.

**TATP:** The results in Fig. 12a show that there is a 26% throughput improvement when using the partitioned models with Houdini. This is mainly attributable to Houdini identifying the best base partition for 82% of TATP's workload that is singled-partitioned (**OP1**, **OP2**). The other 18% first execute a broadcast query on all partitions, thus locking a subset of the partitions is not possible (**OP2**). Subsequent queries in these transactions only access a single partition based on the result of the first query. This also makes it impossible to select the correct base partition for each transaction (**OP1**), since the Houdini cannot know which partition will be needed after the broadcast query. Thus, without the early prepare optimization, all of the other partitions would remain idle (**OP4**), albeit for just a short amount of time. An example of a Markov model for this access pattern is shown in Fig. 10a. Additionally, as shown in Table 4, Houdini disables undo logging for 57.3% of TATP's transactions (**OP3**), but this has negligible impact since the transactions execute only 1-3 queries or are read-only.

The throughput of global models is 4.5% slower on average than the partitioned models due to lock contention in Houdini.

**TPC-C:** This benchmark's results in Fig. 12b show that the "assume single-partition" method performs 6% better than Houdini for small clusters. This is because the likelihood that a transaction is already at the best base partition is greater when there are fewer

94

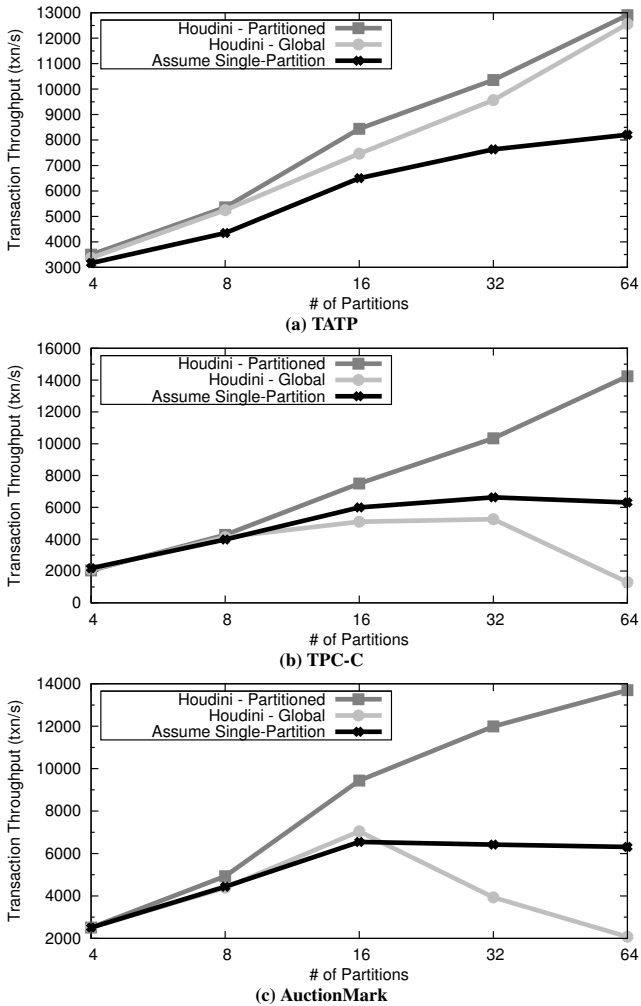

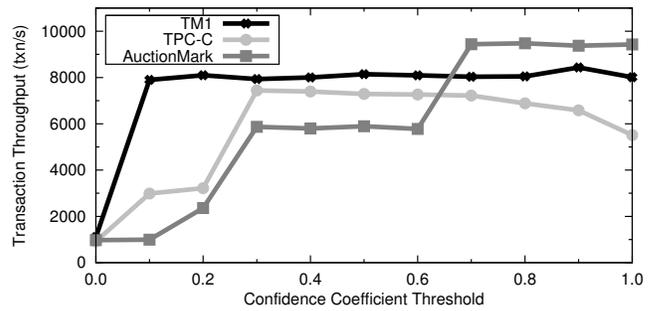

**Figure 13:** Throughput measurements of H-Store under varying estimation confidence coefficient thresholds (Section 4.2).

**Figure 12:** Throughput measurements of H-Store for different execution modes: (1) Houdini with partitioned Markov models; (2) Houdini with global Markov models; and (3) DB2-style transaction redirects and assuming that all partitions are single-partitioned.

partitions (**OP1**). TPC-C's procedures also execute more queries per transaction than the other benchmarks, and thus the estimations take longer to compute. As shown in Table 4, Houdini takes an average of 4 ms to compute estimates for `Delivery` transactions, but these transactions take over 40 ms to execute. The benefit of our techniques therefore is only evident for larger clusters: Houdini's ability to identify not only whether a transaction is distributed or not, but also which partitions it accesses improves throughput by 33.6% (**OP2**). Table 4 also shows that 65.3% of TPC-C's workload is executed without undo logging, most of which are after the transactions have started (**OP3**).

These results also highlight the advantage of model partitioning: the global models' size grows exponentially relative to the number of partitions, thereby increasing the time Houdini needs to traverse the model. The partitioned models also allow Houdini to identify the correct partitions needed for `NewOrder` and `Payment` transactions more often, resulting in fewer aborted transactions. As shown in Fig. 10b, partitioning `Payment`'s models creates almost linear models, which enables Houdini to easily identify when the transaction is distributed (**OP2**).

**AuctionMark:** The results in Fig. 12c show that H-Store achieves an average 47.3% performance improvement when using Houdini with partitioned models for this workload. Like TPC-C, the global models have scalability issues as the size of the cluster increases. AuctionMark mostly benefits from identifying the two partitions the lock for distributed transactions: one for the buyer and one for the seller (**OP2**). As shown in Table 4, Houdini identifies this optimization for 100% of the transactions. The "assume single-partition" strategy does not scale because the transactions do not access the remote partition in the first set of queries (**OP1**), thus they must always be restarted again and lock all of the partitions. Other procedures, such as `GetUserInfo` shown in Fig. 10c, contain conditional branches with separate single-partition and multi-partition paths. Such procedures are ideal for our model partitioning technique, but most of AuctionMark's transactions are short-lived, which means that disabling undo logging (**OP3**) and early prepare optimizations (**OP4**) only provide a modest benefit.

As explained in Section 4.6, we disabled Houdini for the maintenance `CheckWinningBids` requests, as it takes too long process due to the large number of queries (>175) in each transaction. Houdini also does not correctly predict the accessed partitions for 45.0% of `PostAuction` transactions (**OP2**) because their input parameters are large, arbitrary length arrays, which does not work well with our model partitioning technique.

### 6.5 Confidence Sensitivity Analysis

Lastly, we measured how H-Store performs when Houdini uses different confidence coefficient thresholds to select which optimizations to enable. Recall from Section 4.3 that this threshold determines whether a prediction will be included for a transaction based on its confidence coefficient. We executed the benchmarks again in H-Store on a 16-partition cluster and vary the confidence threshold from zero (i.e., all estimations are permitted) to one (i.e., only the most certain estimations are permitted).

As expected, when the threshold is set to zero, the results in Fig. 13 show that all transactions are executed as multi-partition since Houdini predicts that each transaction will always touch all partitions. Once the threshold is >0.06 (i.e., $\frac{1}{16}$), the throughput for TATP remains the same because Houdini correctly identifies which partitions transactions will access (**OP1**, **OP2**) and when they are finished with them (**OP4**). For TPC-C, the throughput reaches a plateau at >0.3 because the number of mis-predicted **OP1** for `NewOrder` transactions is reduced from 10% to 5%, but declines slightly as the threshold approaches one because Houdini no longer selects to disable undo logging as much as it could. In the case of AuctionMark, there are two procedures with conditional branches where Houdini does predict the correct partitions (**OP1**, **OP2**) until the threshold is >0.33 and >0.66.

## 7. RELATED WORK

Modeling workloads using machine learning techniques is a well-known approach for extracting information about a database system [28]. To our knowledge, however, our work is the first to gener-

95

ate intra-transaction execution models (i.e., modeling what queries a transaction executes rather than simply what transactions were executed) and use them to optimize the execution of individual transactions in a parallel database environment. Previous approaches either model workloads based on individual queries or sets of transactions in order to (1) manage resource allocation or (2) estimate future actions of other transactions.

In the former category, the Markov models described in [15, 16] are used to dynamically determine when the workload properties of an application have changed and the database's physical design needs to be updated. The techniques proposed in [10] identify whether a sample database workload is either for an OLTP- or OLAP-style application, and then tunes the system's configuration accordingly. The authors in [11] use decision trees to schedule and allocate resources for long running OLAP queries.

The authors in [22] generate Markov models that estimate the next query that an application will execute based on what query it is currently executing and then pre-fetches that query if there are enough resources available. Similarly, the authors in [9] use Markov models to estimate the next transaction a user will execute based on what transaction the DBMS is executing now. The work described in [27] does use Markov models based on queries much like ours, but their models are designed to identify user sessions across transactional boundaries and to extract additional usage patterns for off-line analysis purposes.

## 8. FUTURE WORK

Representing transactions with Markov models is also applicable to several other research problems in parallel OLTP systems. We plan on extending our models to include additional information about transactions, such as their resource usage and execution times. This information could then be used for admission control or the intelligent scheduling of transactions based on the results of the initial path estimates [11]. For example, the execution states in a model could also include the expected remaining run time for a transaction. By examining the relationships between queries and the procedure parameters, we can discover commutative sets of queries that could then be pre-fetched if the transaction enters some "trigger" state [22]. Similarly, the models could also identify sets of redundant queries in a transaction that could automatically be rewritten and grouped into a smaller batch.

We are currently investigating techniques for the automatic reorganization of on-line H-Store deployments in order to respond to changes in demand and workload skew. We plan on leveraging our models' ability to quickly compare the expected execution paths of transactions with the actual execution properties of the current workload. Such automatic changes include scaling up the number of partitions in the system or repartitioning the database.

## 9. ACKNOWLEDGEMENTS

The authors would like to thank Saurya Velagapudi and Yuri Malitsky for help with this work, Sam Madden for his feedback, and David DeWitt for use of his cluster.

## 10. CONCLUSION

We introduced a new approach for representing the stored procedures of OLTP applications using Markov models to forecast the behavior of transactions. Such models are used to identify when the DBMS can execute a transaction using four different optimizations. From this, we then presented Houdini, a new prediction framework that uses our Markov models to estimate the execution path of future transactions. We described a method for generating these models, as well as how to partition them on certain features to improve their scalability and accuracy. To evaluate our work, we integrated Houdini into the H-Store parallel OLTP system. The results from our experimental analysis show that our models accurately predict the execution paths of 93% of transactions in three OLTP benchmarks. We also demonstrated that our technique has only an average overhead of 5.8%, while increasing the throughput of the system by an average of 41% compared to a naïve approach. These results suggest that predicting transaction properties using Markov models could be useful for any distributed OLTP database. In future work, we will attempt to apply it to real applications and systems.

## 11. REFERENCES

[1] AuctionMark: A Benchmark for High-Performance OLTP Systems. http://hstore.cs.brown.edu/projects/auctionmark.
[2] H-Store: A Next Generation OLTP DBMS. http://hstore.cs.brown.edu.
[3] M. K. Aguilera, A. Merchant, M. Shah, A. Veitch, and C. Karamanolis. Sinfonia: a new paradigm for building scalable distributed systems. In *SOSP*, pages 159–174, 2007.
[4] S. Blott and H. F. Korth. An almost-serial protocol for transaction execution in main-memory database systems. In *VLDB*, pages 706–717, 2002.
[5] S. Chaudhuri and V. Narasayya. Autoadmin "what-if" index analysis utility. *SIGMOD Rec.*, 27(2):367–378, 1998.
[6] J. Coleman and R. Grosman. Unlimited Scale-up of DB2 Using Server-assisted Client Redirect. http://ibm.co/fLR2cH, October 2005.
[7] C. Curino, E. Jones, Y. Zhang, and S. Madden. Schism: a workload-driven approach to database replication and partitioning. *VLDB*, 3:48–57, 2010.
[8] D. J. DeWitt, R. H. Katz, F. Olken, L. D. Shapiro, M. R. Stonebraker, and D. A. Wood. Implementation techniques for main memory database systems. *SIGMOD Rec.*, 14:1–8, June 1984.
[9] N. Du, X. Ye, and J. Wang. Towards workflow-driven database system workload modeling. In *DBTest '09*, pages 1–6, 2009.
[10] S. S. Elnaffar. A methodology for auto-recognizing dbms workloads. In *CASCON*, page 2. IBM Press, 2002.
[11] C. Gupta, A. Mehta, and U. Dayal. PQR: Predicting Query Execution Times for Autonomous Workload Management. In *ICAC*, pages 13–22, 2008.
[12] I. Guyon, S. Gunn, M. Nikravesh, and L. A. Zadeh. *Feature Extraction: Foundations and Applications*. Springer-Verlag, 2006.
[13] M. Hall, E. Frank, G. Holmes, B. Pfahringer, P. Reutemann, and I. H. Witten. The WEKA data mining software: an update. *SIGKDD Explorations Newsletter*, 11:10–18, November 2009.
[14] S. Harizopoulos, D. J. Abadi, S. Madden, and M. Stonebraker. OLTP through the looking glass, and what we found there. In *SIGMOD*, pages 981–992, 2008.
[15] M. Holze and N. Ritter. Towards workload shift detection and prediction for autonomic databases. In *PIKM*, pages 109–116, 2007.
[16] M. Holze and N. Ritter. Autonomic Databases: Detection of Workload Shifts with n-Gram-Models. In *ADBIS*, pages 127–142, 2008.
[17] R. A. Howard. *Dynamic Programming and Markov Processes*. MIT˜Press, 1960.
[18] E. P. Jones, D. J. Abadi, and S. Madden. Low overhead concurrency control for partitioned main memory databases. In *SIGMOD*, pages 603–614, 2010.
[19] S. Kadioglu, Y. Malitsky, M. Sellmann, and K. Tierney. Isac – instance-specific algorithm configuration. In *ECAI*, pages 751–756, 2010.
[20] R. Kallman, H. Kimura, J. Natkins, A. Pavlo, A. Rasin, S. Zdonik, E. P. C. Jones, S. Madden, M. Stonebraker, Y. Zhang, J. Hugg, and D. J. Abadi. H-Store: A High-Performance, Distributed Main Memory Transaction Processing System. *Proc. VLDB Endow.*, 1(2):1496–1499, 2008.
[21] G. Samaras, K. Britton, A. Citron, and C. Mohan. Two-phase commit optimizations and tradeoffs in the commercial environment. In *ICDE*, pages 520–529, 1993.
[22] C. Sapia. PROMISE: Predicting Query Behavior to Enable Predictive Caching Strategies for OLAP Systems. In *DaWaK*, pages 224–233, 2000.
[23] M. Stonebraker, S. Madden, D. J. Abadi, S. Harizopoulos, N. Hachem, and P. Helland. The end of an architectural era: (it's time for a complete rewrite). In *VLDB*, pages 1150–1160, 2007.
[24] Transaction Processing Council. TPC-C Benchmark (Version 5.10). http://www.tpc.org/tpcc/spec/tpcc_current.pdf, 2008.
[25] A. Wolski. TATP Benchmark Description (Version 1.0). http://tatpbenchmark.sourceforge.net, March 2009.
[26] L. Xu, F. Hutter, H. H. Hoos, and K. Leyton-Brown. Satzilla: portfolio-based algorithm selection for sat. *J. Artif. Int. Res.*, 32:565–606, June 2008.
[27] Q. Yao, A. An, and X. Huang. Mining and modeling database user access patterns. In *Foundations of Intelligent Systems*, volume 4203 of *Lecture Notes in Computer Science*, pages 493–503. 2006.
[28] P. S. Yu, M.-S. Chen, H.-U. Heiss, and S. Lee. On workload characterization of relational database environments. *IEEE Trans. Softw. Eng.*, 18(4):347–355.